\documentclass[prl,twocolumn]{revtex4-1}
\usepackage{graphicx}
\usepackage{bm}
\usepackage{textcase}
\usepackage{natbib}
\usepackage{url}
\usepackage{amsmath}
\usepackage{amssymb}
\usepackage{float}
\usepackage{setspace}

\usepackage{lineno} \linenumbers*[1]
\begin{document}

\title{von K\'arm\'an energy decay and heating of protons and electrons in a kinetic turbulent plasma}

\author{P. Wu$^1$}
\email{penny@udel.edu}
\author{M. Wan$^1$}
\author{W. H. Matthaeus$^1$}
\email{whm@udel.edu}
\author{M. A. Shay$^1$}
\author{M. Swisdak$^2$}
\affiliation{$^1$Department of Physics and Astronomy, Bartol Research Institute, University of Delaware,  Newark, Delaware, 19716\\
$^2$Institute for Research in Electronics and Applied Physics, University of Maryland, College Park, Maryland, USA}

\date{\today}
\begin{abstract}
Decay in time of undriven weakly collisional kinetic plasma turbulence in systems large compared to the ion kinetic scales is investigated using fully electromagnetic particle-in-cell simulations initiated with transverse flow and magnetic disturbances, constant density, and a strong guide field. The observed energy decay is consistent with the von K\'arm\'an hypothesis of similarity decay, in a formulation adapted to magnetohydrodyamics (MHD). Kinetic dissipation occurs at small scales, but the overall rate is apparently controlled by large scale dynamics. At small turbulence amplitude the electrons are preferentially heated. At larger amplitudes proton heating is the dominant effect. In the solar wind and corona the protons are typically hotter, suggesting that these natural systems are in large amplitude turbulence regime. 
\end{abstract}
\maketitle

In turbulence theory,  the standard cascade scenario begins with energy supplied at a large (outer) scale, which transfers by a series of approximately local-in-scale nonlinear interaction to smaller (inner) scales where it is dissipated by nonideal microscopic mechanisms. In hydrodynamics this picture is well studied and widely accepted, with energy decay assumed to be independent of viscosity, leading to the von K\'arm\'an-Howarth decay law in which dissipation rates are controlled by dynamics at the outer scale. Turbulence and cascade are also invoked in numerous discussions of dynamics and heating in space and astrophysical plasmas such as the solar corona \cite{Parker72,EinaudiEA96,MattEA99-ch,VerdiniEA10-accn}, the solar wind \cite{Coleman68,TuEA84,Hollweg86,MattEA99-swh}, and the interstellar medium \cite{ArmstrongEA81,MacLow1999,ElmegreenScalo2004a}. Current research on the solar wind often focuses on power law inertial range cascades and microscopic dissipation processes. However the basic underpinnings of the plasma turbulence picture rest on the von K\'arm\'an-Howarth decay conjecture, which has not been directly evaluated for a collisonless magnetized plasma. The extension of this conjecture to plasma dynamics necessarily involves causal detachment of the cascade rate from the dissipation mechanisms. However a baseline question, independent of specific mechanisms, remains as to which microscopic reservoir of internal energy -- protons or electrons -- is the ultimate repository of energy received from the cascade. This letter examines these two questions in a low collisionality plasma: von K\'arm\'an-Howarth energy decay, and heating of protons and electrons. Similarity decay for this kinetic plasma is found, in essentially the form expected for magnetohydrodynamics, while higher amplitude turbulence favors dissipation by protons. This confirms and extends basic principles of turbulence theory to a growing list of applications in space and astrophysical plasmas. 

Similarity decay of energy in hydrodynamics was suggested by Taylor\cite{Taylor35}, and made precise by von K\'arm\'an and Howarth \cite{KarmanHowarth38} who introduced the notion of self-preservation of the functional form of the two point velocity correlation during the decay of isotropic turbulence. Conditions for consistency require that energy ($u^2$) decays as $du^2/dt = -\alpha u^3/L$ while a characteristic length $L$ evolves as $dL/dt = \beta u$, for time $t$, constants $\alpha $ and $\beta$, and similarity variables $u$ (characteristic flow velocity) and $L$ (characteristic eddy size). This familiar formulation has numerous implications for turbulence theory, including ensuring that the dissipation rate is independent of viscosity as required for derivation of the exact third order law\cite{MoninYaglom}. Extensions to energy decay in magnetohydrodynamics (MHD) is often based on dimensional analysis, which provides physically plausible, but non-unique formulations (see e.g., \cite{Biskamp2003,HossainEA95,Matthaeus1996JPP}). When  based on the self-preservation principle, MHD similarity decay involves two Els\"asser energies $Z_+^2$ and $Z_-^2$ and two similarity length scales $L_+$ and $L_-$ \citep[]{Wan2012JFM}. This formulation is based on two conservation laws (energy and cross helicity) in incompressible single fluid MHD, and therefore while it is more complex than hydrodynamics, very little of the richness of kinetic plasma behavior is captured. Its applicability for plasma turbulence might therefore be deemed questionable. On the other hand, there are plausible expectations that MHD is a good description of kinetic plasma dynamics at low frequencies and long wavelengths, especially in directions perpendicular to the magnetic field, e.g., in the solar wind \cite{TuMarsch95}. We see no compelling reason to reject this argument, but even if true, this does not imply that MHD similarity decay is obtained in the plasma case if kinetic effects control dissipation. Therefore we inquire here whether energy decay in a kinetic plasma is consistent with the MHD similarity principle, proceeding numerically, employing a electromagnetic Particle-in-Cell (PIC) method. 

As a step towards plasma behavior, consider the constant density incompressible MHD equation, written in terms of solenoidal velocity ${\bf v}$ and magnetic field ${\bf b}$ in Alfv\'en speed units, pressure $p$, viscosity $\nu$ and resistivity $\mu$. The model includes, a momentum equation $\frac{\partial {\bf v}}{\partial t} + {\bf v} \cdot \nabla {\bf v} = -\nabla p + (\nabla \times {\bf b}) \times {\bf b} + \nu \nabla^2 {\bf v}$ and a magnetic induction equation $\frac{\partial {\bf b}}{\partial t} = \nabla \times ({\bf v} \times {\bf b} ) + \mu \nabla^2 {\bf b}$. For ideal ($\mu \sim 0$, $\nu \sim 0$) incompressible MHD the total energy, kinetic plus magnetic $E = \frac12 \langle |{\bf v}|^2 + |{\bf b}|^2 \rangle$, and the cross helicity $H_c =\langle {\bf v} \cdot {\bf b}\rangle$ are conserved. These are equivalent to the Els\"asser energies $Z_\pm^2 = \langle |z_{\pm}|^2\rangle=\langle |{\bf v} \pm {\bf b}|^2\rangle$ (where $z_{\pm}$ is the  Els\"asser variables) which may be viewed as the cascaded quantities in MHD turbulence theory. Based on the two assumptions of finite energy decay at large Reynolds numbers \cite{KarmanHowarth38} and preservation of the functional form of the correlation functions, for MHD one finds four conditions for consistency of the assumption of similarity decay of energy \cite[]{Wan2012JFM}, namely,
\begin{eqnarray}
\frac {dZ_+^2}{dt} =&& -\alpha_+ \frac{Z_+^2 Z_-}{L_+}; 
\hspace{.1in}
\frac {dZ_-^2}{dt} = -\alpha_- \frac{Z_-^2 Z_+}{L_-}\\
\frac {dL_+}{dt} =  && \,\,\,\,\, \beta_+ Z_-;
\hspace{.3in}
\frac {dL_-}{dt} = \,\,\,\,\, \beta_- Z_+
\label{eq:similarity}
\end{eqnarray}
This generalizes the von K\'arm\'an-Howarth result \cite[]{KarmanHowarth38} to fully isotropic MHD or MHD isotropic in a plane transverse to a strong mean field \cite[]{Wan2012JFM}.
 
{\it Simulations.}
To test the hypothesis that a turbulent kinetic plasma might follow von K\'arm\'an-Howarth energy decay in the MHD form, we carry out a set of PIC simulations. We opt for 2.5 dimensional (D) geometry (2D wave vectors, 3D velocity and electromagnetic fields) in order to attain sufficient scale separation, equivalent to large effective Reynolds numbers, typically regarded as a condition for similarity decay \cite{vonKarman1949}. Here scale separation requires that the outer scales $L_{\pm}$ are substantially greater than the dissipative scales, nominally associated with the ion inertial scale $d_{i}$. 

The fully electromagnetic PIC simulations \cite[]{Zeiler2002} employed here solve the kinetic equations using super-particles that respond to the Lorentz force, coupled to Maxwell's equations. The simulation is normalized to reference parameters: density $n_{r}=1$, magnetic field $B_{r}=1$, and mass (ion mass) $m_{i}=1$; as well as derived (from $n_{r}, B_{r}, m_{i}$) parameters, the ion inertial length $d_{i}$, the ion cyclotron time $\Omega_{i}^{-1}$, the Alfv\'en speed $v_{Ar}$, and the temperature scale $T_{r}=m_{i}v_{Ar}^{2}$. For simplicity, in the following, we will employ dimensionless units unless otherwise specified.

A summary of run parameters is given in Table \ref{table1}. Run 2 (the reference run) is in a $(25.6 d_i)^2$ box, with $2048^2$ grid points. Initially, there are 300 particles per cell with uniform density $n_{0}=1$. The initial temperature of ions and electrons is $T_{0} = 1.25$ (normalized to $m_i v^2_{Ar}$). The Debye length, $\lambda_D=0.05$, is more than $4 \times$ grid scale. The electron mass and speed of light are $m_{e}=0.04$, and $c=30$, respectively. The time step is $\delta t = 0.0025$. A strong out-of-plane guide field $B_z = 5$ is imposed to reduce compressibility, which gives the system an Alfv\'en speed $v_{A}\sim5v_{Ar}$, an ion cyclotron time $\omega_{ci}^{-1}\sim0.2\Omega_{i}^{-1}$ and a total plasma beta $\beta=0.2$. Initial turbulence is solenoidal velocity, transverse to ${\bf B}_z$ (``Alfv\'en mode'') with unit total fluctuation energy, controlled cross helicity $H_c$, and controlled Alfv\'en ratio $r_{A}=E_{v}/E_{B}=1.0$; see Table. We initialize a Fourier spectrum: $E(k)  \sim  [1 + (k/k_{0})^{8/3}]^{-1}$, for wavenumbers $k = [2, 4]2\pi/25.6$ with $k_0 = 6\times 2\pi/25.6$.  

The selected Runs differ from the reference Run 2 by the highlighted bold parameters (Table \ref{table1}). In run 4, the Alfv\'en ratio is $r_{A}=1.0$ as in Run 2 but the in-plane fluctuating magnetic and velocity are doubled. Run 5 only differs from Run 2 in system size, being $(51.2 d_i)^2$ ($4096^2$ grid points), therefore the corresponding wavenumbers are $k = [2, 4]2\pi/51.2$ with $k_0 = 6\times 2\pi/51.2$. 

\begin{table*} [!]
\caption{Runs: differences from run 2 are highlighted in bold.\label{table1}. The nonlinear time $t_{nl}=(L_{+}(0)+L_{-}(0))/2/z(0)$ (where $z(0)=\sqrt{<z_{+}(0)^2>+<z_{-}(0)^2>}$) is listed in the unit of the system cyclotron time $\omega_{ci}^{-1}$.}
 \begin{ruledtabular}
 \begin{center}
 \begin{tabular}{c|c|c|c|c|c|c|c|c|c|c|c|c}
Runs & 1 & 2 & 3 & 4 & 5 & 6 & 7 & 8 & 9 & 10 & 11 & 12\\
\cline{1-13}
$Z_{0}^2$ & 2  & 2 & 2 & \textbf{8} & 2 & \textbf{8} & \textbf{4.5} & \textbf{18} & \textbf{8} & 2 & 2 & \textbf{12.5}\\
$r_{A}$ & \textbf{0.2} & 1.0 & \textbf{5.0} & 1.0 & 1.0 & \textbf{0.2} & 1.0 & 1.0 & \textbf{5.0} & 1.0 & 1.0 & 1.0\\
size & 25.6 & 25.6 & 25.6 & 25.6 & \textbf{51.2} & 25.6 & 25.6 & 25.6 & 25.6 & 25.6 & \textbf{102.4} & 25.6\\
$H_{c}$ & 0.0 & 0.0 & 0.0 & 0.0 & 0.0 & 0.0 & 0.0 & 0.0 & 0.0 & \textbf{0.8} & 0.0 & 0.0\\
$t_{nl}$& 19.1 & 14.9 & 11.6 & 7.4 & 29.7 & 9.6 & 9.9 & 5.0 & 5.8 & 12.9 & 59.5 & 6.0 \\
\end{tabular}
\end{center}
 \end{ruledtabular}
\end{table*}

{\it Results.}
To study energy decay we examine the time variation of the Els\"asser energies $Z_+^2(t) $ and $Z_-^2(t)$. At each time $t$ of the analysis we compute the two-point correlation functions for the Els\"{a}sser variables $z_{\pm}$, that is, $R_{\pm}(r) = \langle{\bf  z}_{\pm}({\bf x} ) \cdot {\bf z}_{\pm}({\bf x} + {\bf r}) \rangle$ for spatial average $\langle ...\rangle$ and spatial lag ${\bf r}$. We find the lag values $L_\pm$ that solve $R_{+} (L_+) = 1/{\rm e}$ and $R_{-}(L_-) = 1/{\rm e}$ where ${\rm e}=2.71828...$. This defines the outer scales $L_\pm(t)$ at each time. According to the MHD decay hypothesis, the evolution of  $Z_+^2(t) $ and $Z_-^2(t)$ depends on  $Z_+^2(t) $, $Z_-^2(t)$,  $L_+$, and $L_-$, with the variations due to all other effects relegated to implicit dependence of the von K\'arm\'an constants $\alpha_+$ and $\alpha_-$.

\begin{figure}
\begin{centering}
\includegraphics[scale=.47]{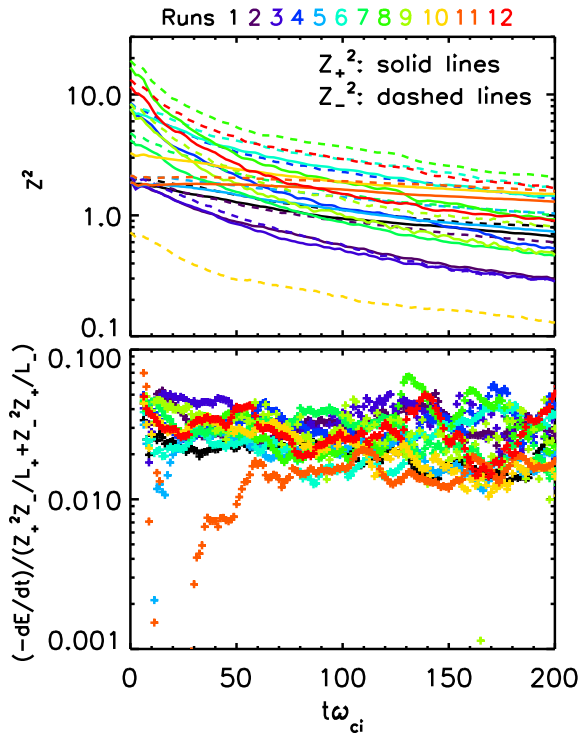}
\caption{(Top) Time history of Els\"asser energies $Z_+^2(t)$ (solid lines) and $Z_-^2(t)$ (dashed lines) and  (bottom) simulated decay rates $dE/dt$ over their respective theoretical expectation $ D_{th} \equiv Z_+^2 Z_- /L_+ + Z_-^2 Z_+/L_-$ for all 12 runs (in 12 colors).}
\label{fig1}
\end{centering}
\end{figure}

Proceeding in this manner, Figure \ref{fig1}, top panel, shows the time history of the Els\"asser energies $Z_+^2(t)$ and $Z_-^2(t)$ 
for the 12 runs listed in Table I. The emphasis in this illustration is not the specific behavior of any individual run, but rather the general trend and time scales of energy decay, and the substantial spread in values in the different runs. To compare with the MHD similarity decay, we examine the decay rate of the energies by numerical evaluation of $dZ_+^2(t)/dt$ and $dZ_-^2(t)/dt$ from each run and combining them to obtain the empirical value of the sum $dZ^2/dt$. 
This is done for clarity of presentation, but  $Z_+^2(t)$ and $Z_-^2(t)$ separately
have been found to equally well follow similarity decay.
In MHD the theoretical expectation is that the decay rate, assuming for simplicity that $\alpha_+ = \alpha_-$, is proportional to $ D_{th} \equiv Z_+^2 Z_- /L_+ + Z_-^2 Z_+/L_-$. Normalizing the empirical decay rate by the theoretical expectation $D_{th}$ permits evaluation of the similarity hypothesis. The result of this normalization is shown in the second panel of Fig. \ref{fig1}, where all runs reach the decay law after one eddy turn over time ($t_{nl}$ in Table \ref{table1}). 
Note that run 11, having the largest system size, requires a longer time to attain a fully developed state, as expected since $t_{nl}$ increases with energy-containing scale. The result is encouraging with regard to the accuracy of similarity decay as it applies to an ensemble of runs, as some case-to-case variability is always expected in turbulence. It is also likely that the values of $\alpha_\pm$, both expected to be $O(1)$, are generally unequal, a possibility we defer to a later time. The level of variability seen here is comparable to analogous variability seen in similarity decay in hydrodynamics \cite{MoninYaglom}, in electron fluids \cite{RodgersEA10} and in MHD runs \cite{HossainEA95}. Therefore we can conclude that the von K\'arm\'an-MHD similarity theory provides a reasonable baseline description of the scaling in time of the energy decay.

\begin{figure}
\begin{centering}
\includegraphics[scale=.4]{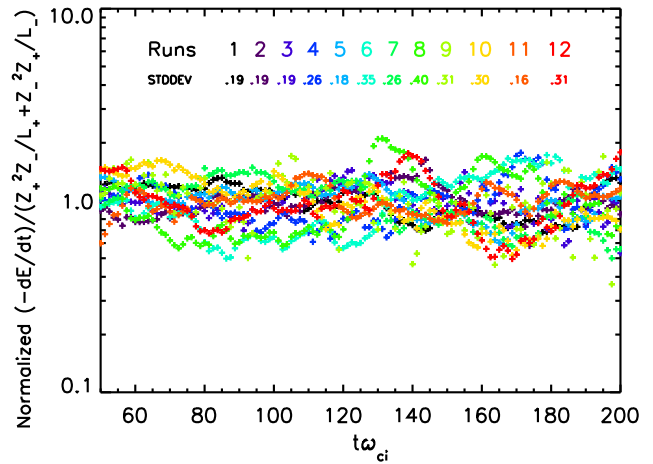}
\caption{Normalized (by the average value of each run's effective decay constant $\alpha^* =(dZ^2/dt)/D_{th}$) decay rates $dE/dt$ over theoretical expectations $ D_{th} \equiv Z_+^2 Z_- /L_+ + Z_-^2 Z_+/L_-$. The standard deviation about unity of the normalized decay rate is listed for each run.}
\label{fig2}
\end{centering}
\end{figure}

This examination may be taken a step further by optimizing the von K\'arm\'an constants for each run, which then takes into account the variation of  physical parameters such as Alfv\'en ratio, Reynolds number, etc., that are not represented explicitly in the similarity theory. In principle this could be a massive effort, requiring many times the number of runs we exhibit here. However a simple way to proceed is to normalize each run by the average value of its effective decay constant $\alpha^* =(dZ^2/dt)/D_{th}$ where each quantity is computed from the run for all times after $\Omega_{ci} t = 50$. In effect this eliminates variability due to possible weak dependence of $\alpha_\pm$ on other parameters. The result of this analysis on the same 12 runs is shown in Fig. \ref{fig2}.  The time series now have an average values of $1$ by construction, but it is also apparent that (i) the variability of the decay rates is reduced; and (ii) the series are all visibly stationary and without trend. We conclude that after a transient startup phase, the ensemble of kinetic plasma turbulence runs exhibits energy decay that is consistent with the MHD extension of the von K\'arm\'an similarity decay hypothesis.

\begin{figure}
\begin{centering}
\includegraphics[scale=.35]{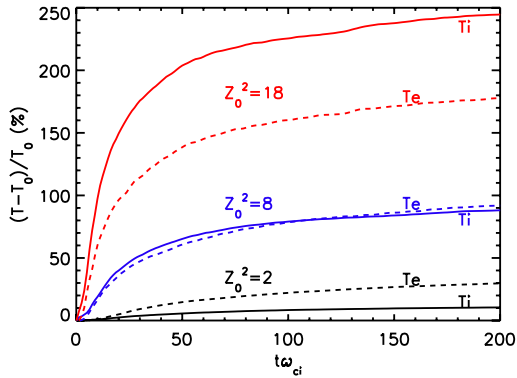}
\caption{Time evolution of ion (solid) and electron (dash) temperatures ($T_{i}$, $T_{e}$) increments overs the initial value $T_{0}$ for runs 2 (black, low initial Els\"asser energy $Z_{0}^2=2$), 4 (blue, medium initial Els\"asser energy $Z_{0}^2=8$), and 8 (red, high initial Els\"asser energy $Z_{0}^2=18$).}
\label{fig3}
\end{centering}
\end{figure}

\begin{figure}
\begin{centering}
\includegraphics[scale=.35]{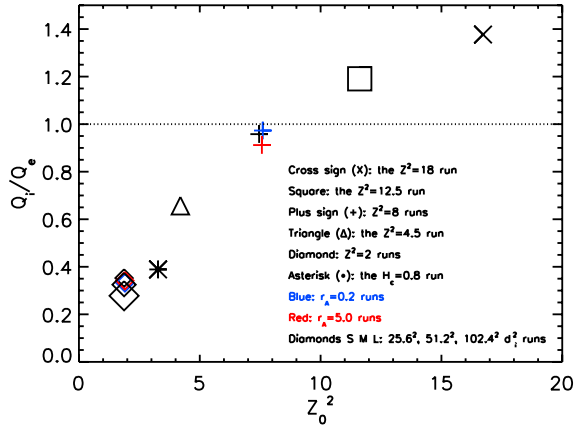}
\caption{Time averaged heat functions $Q_{i}/Q_{e}=\Delta T_{i}/\Delta T_{e}$, where $\Delta T$ is the temperature change from $t=0$ to $t=200 \omega_{ci}^{-1}$, for protons and electrons for all 12 runs.}
\label{fig4}
\end{centering}
\end{figure}

In MHD the loss of energy from fluid scales is due to viscosity and resistivity and known dissipation functions. However for a low collisionality 
plasma, the dissipation function is unknown, and dissipation may involve many processes. This is an active area of interest in space physics and astrophysics \cite{Leamon1998APJ,Gary2009,Bale2005, Matteini2012, Schekochihin2009, ServidioEA12}. 
The basic question of 
how dissipated energy is partitioned between protons 
and electrons \cite{Cranmer2009} is readily addressed 
using the set of PIC runs employed above, 
for which $T_i=T_e$ initially. 
Fig.~\ref{fig3} shows the temperatures evolution for three runs: 
For the low initial energy case the $T_e$ 
increases more than $T_i$. 
For the intermediate energy case the increases in 
$T_i$ and $T_e$ are almost equal. 
Finally, for the strongest turbulence case, 
the proton heating is greater. 
A summary of this result 
is given in Fig. \ref{fig4} which shows $Q_i/Q_e$, 
the ratio of time averaged heat functions ($Q_{i}/Q_{e}=\Delta T_{i}/\Delta T_{e}$ $\propto$ temperature change) 
for protons and electrons for all cases. 
It is apparent that there is a systematic 
increase of proton heating relative to electron heating as turbulence level is increased.
 
{\it Conclusions.}
Adopting an empirical approach based on PIC runs, we have demonstrated that a low collisionality kinetic proton-electron plasma experiences decay of total fluid scale fluctuation energy according to a von K\'arm\'an similarity law. The specific decay law that we employed is derived for incompressible MHD based on the assumption that the shape of two point correlation functions remains unchanged during decay, which implies the decay laws given in Eqs. (1--2). The approximate validity of this approach provides a strong basis for treatments of plasma turbulence, as it does for hydrodynamics \cite[]{MoninYaglom}. This principle assumes that energy decay at outer scales does not depend on details of the microscopic kinetic dissipation processes. It leaves open the question of the proportion of dissipated energy that goes into the electrons and protons. We employed the same set of runs to address this question, and arrived at the potentially important conclusion that more energy goes into the electron heating for low turbulence energies, and more into protons at high initial turbulence levels. The crossover value of the turbulence amplitude occurs when the initial turbulence is such that $\delta b/B_0 \approx 2/5$.

There has been somewhat of a puzzle regarding solar wind and coronal heating, in that observed proton temperatures are usually found to be greater than electron temperatures, but familiar mechanisms such as Landau damping mainly heat electrons. The present work clarifies this situation in an agreeable way, without any contradiction of prior ideas. At small amplitudes the present result is consistent with linear Vlasov theory,  finding dominant heating of electrons. Increased proton heating for stronger turbulence strongly suggests an involvement of coherent structures in kinetic processes, also reported in various recent plasma simulation studies \cite[]{ParasharEA09,Markovskii2010,WanEA12,Karimabadi2013,WuEA13}.

A simple understanding is provided by appealing to the structure of the Kolmogorov refined similarity hypothesis, from which we expect that
\begin{equation}
|\delta_r z|^3 \sim \epsilon_{r} r \label{krsh}
\end{equation}
where $\epsilon_r$ is the total dissipation in a sphere of radius $r$ at position $\bf x$ , and $\delta_r z = \bf e_{r}\cdot ({\bf z}({\bf x}+{\bf r} ) -{\bf z}({\bf x}))$ is the longitudinal increment of an Els\"asser field at spatial lag $r$ (where $\bf e_{r}$ is the unit vector in $\bf r$ direction). Stronger turbulence will have larger $\epsilon_{r}$ and therefore larger increments $\delta_{r}z$. This corresponds to stronger gradients, in particular at coherent structures such as current sheets. However, it is established \cite{DmitrukEA04} that protons interact strongly with currents sheets having a typical scale of the ion inertial length. Stronger current sheets at this scale will open up more channels for kinetic couplings and instabilities. Having more such channels, the protons will be heated more. At lower turbulence levels, there are less couplings at ion scales, and the energy cascade more readily passes through the proton scales without producing dissipation. In that case more of the energy arrives at electron scales where damping will occur. The same basic physical argument has been previously stated \cite{MatthaeusEA09} in regard the variation of the Taylor microscale, and the dependence of sub-ion-inertial scale spectral slope on cascade rate\cite{Smith2006APJ}. The idea that additional ion dissipation channels open up at larger turbulence level/cascade rate is, as far as we know, the only explanation that has been offered for these observed phenomena. Here, the same rationale provides a preliminary explanation for the result that stronger cascades preferentially heats protons. Further work is needed to support and explain this hypothesis, and if it is correct, it may lead to further studies in turbulence theory, plasma processes, simulations and observations. Some results will be forthcoming from these efforts, while we also await attempts to extend these findings.

{\it Acknowledgements.}
This work 
is supported in part by NSF (AGS-1063439, ATM-0645271, 
AGS-1156094), and by NASA 
(NNX09AG31G, NNX11AJ44G, NNX13AD72G, MMS-IDS NNX08A083G, 
MMS Theory and modeling Team, ISIS/Solar Probe Plus). 
Simulations were done on NCAR Yellowstone supercomputers. 



\begin{thebibliography}{10}

\bibitem{Parker72}
E.~N. Parker.
\newblock {\em Astrophys.\ J.}, 174:499--510, 1972.

\bibitem{EinaudiEA96}
G.~Einaudi, M.~Velli, H.~Politano, and A.~Pouquet.
\newblock {\em Astrophys.\ J.}, 457:L113, 1996.

\bibitem{MattEA99-ch}
W.~H. Matthaeus, G.~P. Zank, S.~Oughton, D.~J. Mullan, and P.~Dmitruk.
\newblock {\em Astrophys.\ J.}, 523:L93--L96, 1999.

\bibitem{VerdiniEA10-accn}
A.~Verdini, M.~Velli, W.~H. Matthaeus, S.~Oughton, and P.~Dmitruk.
\newblock {\em Astrophys.\ J.}, 708:L116--L120, 2010.

\bibitem{Coleman68}
P.~J. Coleman.
\newblock {\em Astrophys.\ J.}, 153:371--388, 1968.

\bibitem{TuEA84}
C.-{Y}. Tu, Zu-{Y}in Pu, and Feng-{S}i Wei.
\newblock {\em J.\ Geophys.\ Res.}, 89:9695--9702, 1984.

\bibitem{Hollweg86}
J.~V. Hollweg.
\newblock {\em J.\ Geophys.\ Res.}, 91:4111, 1986.

\bibitem{MattEA99-swh}
W.~H. Matthaeus, G.~P. Zank, C.~W. Smith, and S.~Oughton.
\newblock {\em Phys.\ Rev.\ Lett.}, 82:3444--3447, 1999.

\bibitem{ArmstrongEA81}
J.~Armstrong, J.M. Cordes, and B.J. Rickett.
\newblock {\em Nature}, 291:561, 1981.

\bibitem{MacLow1999}
M.~Mac~Low.
\newblock {\em ApJ}, 524:169--178, 1999.

\bibitem{ElmegreenScalo2004a}
B.~G. Elmegreen and J.~Scalo.
\newblock {\em Annu. Rev. Astron. Astrophys.}, 42:211--273, 2004.

\bibitem{Taylor35}
G.~I. Taylor.
\newblock {\em Proc.\ Roy.\ Soc.\ Lond.\ A}, 151:421--444, 1935.

\bibitem{KarmanHowarth38}
T.~de~K{\'a}rm{\'a}n and L.~Howarth.
\newblock {\em Proc. Roy. Soc. London Ser. A}, 164:192--215, 1938.

\bibitem{Biskamp2003}
D.~Biskamp.
\newblock Cambridge U. Press, Cambridge, UK, 2003.

\bibitem{HossainEA95}
M.~Hossain, P.~C. Gray, Jr. D.~H.~Pontius,
W.~H. Matthaeus \& S. Oughton,
\newblock {\em Phys. Fluids}, 7, 2886, 1995.

\bibitem{Matthaeus1996JPP}
W.~H. Matthaeus, G.~P. Zank, and S.~Oughton.
\newblock {\em J. Plasma Physics}, 56 (3):659--675, 1996.

\bibitem{Wan2012JFM}
M.~Wan, S.~Oughton, S.~Servidio, and W.H. Matthaeus.
\newblock {\em Journal of Fluid Mechanics}, 697:296, 2012.

\bibitem{TuMarsch95}
C.-{Y}. Tu and E.~Marsch.
\newblock {\em Space Sci.\ Rev.}, 73:1--210, 1995.

\bibitem{vonKarman1949}
Theodore von K\'arm\'an and C.~C. Lin.
\newblock {\em Rev. Mod. Phys.}, 21:516--519, Jul 1949.

\bibitem{Zeiler2002}
A.~Zeiler, D.~Biskamp, J.~F. Drake, B.~N. Rogers, M.~A. Shay, and M.~Sholer.
\newblock {\em J. Geophys. Res.}, 107:1230, 2002.

\bibitem{MoninYaglom}
A.~S. Monin and A.~M. Yaglom.
\newblock MIT Press, Cambridge, Mass., 1971, 1975.

\bibitem{RodgersEA10}
D.~J. Rodgers, W.~H. Matthaeus, T.~B. Mitchell \&
 D.~C. Montgomery.
\newblock {\em Phys. Rev. Lett.}, 105, 234501, 2010.

\bibitem{Leamon1998APJ}
R.~J. Leamon, W.~H. Matthaeus, and C.~W. Smith.
\newblock {\em ApJ}, 507:L181--L184, 1998.

\bibitem{Gary2009}
S.~P. Gary and C.~W. Smith.
\newblock {\em J.\ Geophys.\ Res.}, 114, 2009.

\bibitem{Bale2005}
S.~D. Bale, P.~J. Kellogg, F.~S. Mozer, T.~S. Horbury, and H.~Reme.
\newblock {\em Phys. Rev. Lett.}, 94:215002--1, 2005.

\bibitem{Matteini2012}
L.~Matteini, P.~Hellinger, S.~Landi, P.~M. Travnicek, and M.~Velli.
\newblock {\em Space Sci. Rev.}, 172:373--396, 2012.

\bibitem{Schekochihin2009}
A.A.~Schekochihin, S.C.~Cowley, W.~Dorland, G.W.~Hammett, G.G.~Howes, E.~Quataert
\&  T.~Tatsuno.
\newblock {\em Astrophys. J. Supp.}, 182(1):310, 2009.

\bibitem{ServidioEA12}
S.~Servidio, F.~Valentini, F.~Califano, and P.~Veltri.
\newblock {\em Phys.\ Rev.\ Lett.}, 108:045001, 2012.

\bibitem{Cranmer2009}
S.~R. Cranmer, W.~H. Matthaeus, B.~A. Breech, and J.~C. Kasper.
\newblock {\em Astrophys. J.}, 702:1064--1614, 2009.

\bibitem{ParasharEA09}
T.~N. Parashar, M.~A. Shay, P.~A. Cassak, and W.~H. Matthaeus.
\newblock {\em Phys.\ Plasmas}, 16(3):032310, 2009.

\bibitem{Markovskii2010}
S.~A. Markovskii and B.~J. Vasquez.
\newblock {\em Phys. Plasmas}, 17:112902, 2010.

\bibitem{WanEA12}
M.~{Wan}, W.~H. {Matthaeus}, H.~{Karimabadi}, V.~{Roytershteyn}, M.~{Shay},
  P.~{Wu}, W.~{Daughton}, B.~{Loring} \& S.~C. {Chapman}.
\newblock {\em Phys. Rev. Lett.}, 109, 195001 (2012).

\bibitem{Karimabadi2013}
H.~Karimabadi, V.~Roytershteyn, M.~Wan, W.~H. Matthaeus, P.~Wu W.~Daughton,
  M.~A. Shay, B.~Loring, J.~Borovsky, E.~Leonardis, S.~C. Chapman \& T.~K.~M.
  Nakamura,
\newblock {\em Phys. Plasmas}, 012303, 2013.

\bibitem{WuEA13}
P.~{Wu}, S.~{Perri}, K.~{Osman}, M.~{Wan}, W.~H. {Matthaeus}, M.~A. {Shay},
  M.~L. {Goldstein}, H.~{Karimabadi}, and S.~{Chapman}.
\newblock {\em Astrophys. J. Lett.}, 763, L30 (2013).

\bibitem{DmitrukEA04}
P.~{Dmitruk}, W.~H. {Matthaeus}, and L.~J. {Lanzerotti}.
\newblock {\em Geophys. Res. Lett.}, 31:21805, November 2004.

\bibitem{MatthaeusEA09}
W.~H. {Matthaeus}, S.~{Oughton}, and Y.~{Zhou}.
\newblock {\em Phys. Rev. E}, 79(3):035401, March 2009.

\bibitem{Smith2006APJ}
C.~W. Smith, K.~Hamilton, and B.~J. Vasquez.
\newblock {\em Astrophys. J.}, 645:L85--L88, 2006.

\end{thebibliography}

\end{document}